\documentclass[10pt,English]{iopart}
\usepackage{graphicx}
\usepackage{epstopdf}
\usepackage{float}
\usepackage{color}
\usepackage{hyperref}
\begin{document}

\title[On reversible,  endoreversible and irreversible heat device cycles vs Carnot cycle]{On reversible, endoreversible, and irreversible heat device cycles versus Carnot cycle: A pedagogical approach to account for losses}

\author{Gonzalez-Ayala J\textsuperscript{[1]}, Angulo-Brown F\textsuperscript{[2]}, Calvo Hern\'andez\textsuperscript{[1]} A and Velasco S\textsuperscript{[1]}}

\address{\textsl{\textsuperscript{[1]} Departamento de F\'isica aplicada, Universidad de Salamanca, 37008 Salamanca, Spain\\
\textsuperscript{[2]} Departamento de F\'isica, Escuela Superior de F\'isica y Matem\'aticas, Instituto Polit\'ecnico Nacional, Edif. No. 9, U.P. Zacatenco, 07738, M\'exico D. F., M\'exico}}

\begin{abstract}
In this work we analyze the deviations of reversible cycles (for both heat engines and refrigerators) from the corresponding Carnot cycle operating between the same extreme temperatures and deviations of irreversible cycles from its corresponding reversible realization while putting emphasis on the corresponding losses. The endoreversible models fit in the proposed framework. Two suitable loss factors, which don't need the explicit calculation of entropy variations, are introduced.
The behavior of these factors and their interplay allow for a clear and pedagogical visualization of where external and internal irreversibilities  are located and their intensities in terms of the main variables describing the cycle. 
The analysis could be used as  a starting point for more advanced studies on modeling and optimization of real devices and installations.
\end{abstract} 
\textit{This is the preprint of the published verison available at \href{https://doi.org/10.1088/0143-0807/37/4/045103}{10.1088/0143-0807/37/4/045103}}
\ioptwocol
\section{Introduction}\label{s1}
It is known from the maximum work theorem~\cite{zemansky97,cengel02,callen85} that for all processes, leading a thermodynamical system from an initial to a final state, the delivery of work is maximum for a reversible process. Moreover, from Carnot theorem and its equivalency with second law of thermodynamics it is known that the efficiency of an energy converter operating between two external reservoirs is given by the Carnot efficiency, for heat engines, and the Carnot coefficient of performance (COP) for refrigerators. We stress that reversibility along any thermodynamic process requires a succession of infinite equilibrium states, thus implying infinite time of realization. This is not the case of heat engines in real life, where we use heat engines working with a non-zero net power output operating at finite times.
In this way, it has been long known that the actual performance of any heat device is accompanied by losses in regard to
 a reversible realization, meanwhile the reversible realization will display a decrement in its performance with respect to the Carnot realization. 

Below, two statements over a generic non-isothermal energy converter will be used here to account for two classes of losses.
We consider both heat engines and refrigerators devices and the focus is on the losses of each device for the specific job it was designed: to produce work for a heat engine and to extract cooling heat for a refrigerator. To get this and from the perspective of an introductory thermodynamics, we introduce in each case two suitable loss factors, which don't need the explicit calculation of entropy variations, a subtle and sometimes difficult task. The endoreversible systems composed by an inner reversible subsystem  coupled with two external reservoirs by an irreversible interaction are included in a natural way in the proposed framework.

 These factors, closely related with the second-law efficiency~\cite{bejan06,marcella92}, allow for a clear visualization of the behavior of losses, their nature and intensity in terms of the main variables describing the cycle as well as of the resulting parametric power versus efficiency curves. Thus, it can guide a first approach to the design and optimization of real heat devices, an issue of primary interest due to the energy shortage~\cite{durmayaz04}.  Besides, the results can be instructive in order to stressing how the imperfections of any heat converter, mainly irreversibilities coming from finite-time and finite-size, provoke that the heat  input in a heat engine is not fully available as work output, a situation we face also every day in real life.

The paper is organized as follow: In Sect.~2 we define the two loss factors on thermodynamics basis. In order to show how the loss factors work and their influence on the main energetic magnitudes we present two representative and pedagogical cycles: a Brayton-like cycle in Sect.~3 and a Carnot-like irreversible heat engine in Sect.~4. In Sect.~5 the analysis is extended to refrigeration cycles in order to show its unified validity with heat devices. Finally, some specific remarks on the loss factors are presented in Sect.~6.

\section{Heat engine cycles.}\label{s2}

\begin{figure}[t]\centering
\includegraphics[width=\linewidth]{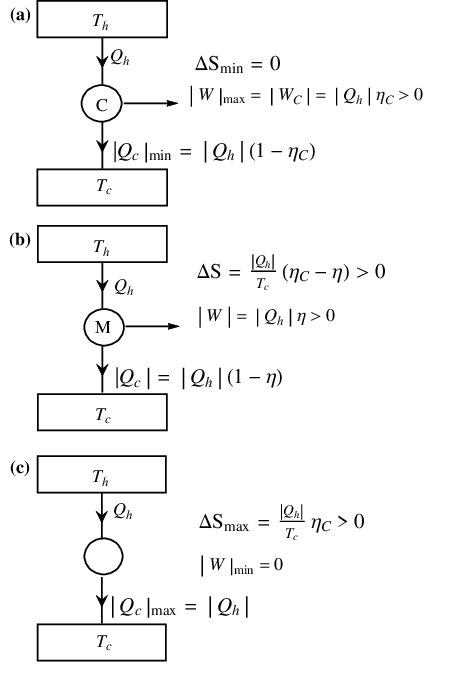}
\caption{Schematic diagram of a Carnot cycle (a), a generic heat engine (b) and a direct heat leak (c) between two external reservoirs at temperatures $T_h$ and $T_c$ (see text). Work output decreases from the maximum possible value $W_C$ in (a) to its minimum $W=0$ value in (c).}
\label{fig.1}
\end{figure}
Usually the second law of Thermodynamics for closed systems is expressed as the increasing entropy principle $\Delta S\ge 0$~\cite{zemansky97,cengel02,callen85,bejan06,marcella92}, where the equality applies to reversible processes. When applied to a heat engine performing a cyclic process it states that
 the best conversion of an amount of heat input $Q_h\geq0$ in work output $W\leq 0$ is trough a Carnot heat engine. For such a conversion, see Fig.~1a, the efficiency is $\eta_C=1-\tau$, where $\tau\equiv T_c/T_h\leq 1$ is the ratio of the temperature of the cold thermal bath $T_c$ to the temperature of the hot thermal bath $T_h$; the work output $|W|_{max}=|Q_h|\eta_C\equiv |W_C|$ is thus the maximum possible work delivered from a fixed amount of heat $Q_h$. Obviously, the variation of the overall entropy  is $\Delta S=-\frac{|Q_h|}{T_h}+\frac{|Q_c|}{T_c}=0$ since the inherent reversibility of the Carnot cycle and no lost work is generated $|W_{lost}|=T_c\Delta S=0$. On the opposite side we can imagine a situation where the same heat $Q_h$ is entirely dissipated in the cold thermal bath without the realization of work output (see Fig.~1c). In these circumstances $\Delta S=-\frac{|Q_h|}{T_h}+\frac{|Q_h|}{T_c}=\frac{|Q_h|}{T_c}\eta_c\equiv \Delta S_{max}$ is the maximum possible value of the entropy variation and, as a consequence,
$|W_{lost}|=T_c\Delta S_{max}$ amounts  the maximum possible value that becomes unavailable to do work. In between these two limit situations we consider a generic heat engine $M$ working between the same external heat reservoirs $T_h$ and $T_c$ and with the same heat input $Q_h$ (see Fig.~1b). 
 Now, according with the first law for any cyclic process ($\Delta U=0$) $|W|=|Q_h|-|Q_c|=|Q_h|(1-\frac{|Q_c|}{|Q_h|})\equiv 
|Q_h|\eta$, where $\eta$ is 
the first-law conversion efficiency (i.e. the fraction of heat input converted in work output) and $\Delta S=-\frac{|Q_h|}{T_h}+\frac{|Q_c|}{T_c}=\frac{|Q_h|}{T_c}(\eta_c-\eta)$. In order to account for deviations in work output from the Carnot performance we propose a total loss factor $d$ measuring the lost work given by the heat engine $M$ with respect to the maximum possible work output performed by the Carnot cycle, $W_C$:

\begin{eqnarray}
d=\frac{|W|_{lost}}{|W|_{max}}=\frac{|W|_{max}-|W|}{|W|_{max}}\equiv \frac{|W|_{C}-|W|}{|W|_{C}}= \nonumber\\
=\frac{\eta_C-\eta}{\eta_C}\leq 1
\end{eqnarray}
where the last equality is due to the assumption of the same heat input. 

From this definition we can divide the total losses distinguishing the reversible realization of $M$ from the corresponding Carnot cycle, and the real irreversible performance from the reversible case, respectively, by using the following relation:
\begin{equation}
d=\frac{\eta_C-\eta}{\eta_C}\equiv \frac{\eta_C-\eta_{rev}}{\eta_C}+\frac{\eta_{rev}-\eta}{\eta_{rev}} \left(\frac{\eta_{rev}}{\eta_C}\right) \leq 1
\end{equation}
or
\begin{equation}
d=d_e+\left(\frac{\eta_{rev}}{\eta_C}\right)d_{i}=d_e+d_{i}-d_e d_{i} \label{eq:dHE}
\end{equation}
where
\begin{equation}
d_e=1-\frac{\eta_{rev}}{\eta_C}\leq 1
\end{equation}
and
\begin{equation}
d_{i}=1-\frac{\eta}{\eta_{rev}} \leq 1
\end{equation}

The factor $d_e$ accounts for differences in work output
between the Carnot and the idealized (reversible) cycles while the second one, $d_{i}$, is a measure of the difference between the idealized and actual performances. Note in Eq.~3 that $d=d_e+d_i-d_ed_i$ includes a coupling term $d_ed_i$ as a notable feature.
 
Obviously, if $M$ is a reversible Carnot heat engine working between the same extreme temperatures and with the same heat input $d_e=0$. For any different reversible cycle $\eta_{rev}\leq \eta_{C}$ and $d_e\leq1$. Since $d_e$ involves only the consideration of reversible processes, this loss factor could be interpreted and measured in geometric terms: for the same heat input in the Carnot and in the idealized heat engines, $d_e=\frac{\eta_C-\eta_{rev}}{\eta_C}=\frac{|W_{C}|-|W_{rev}|}{|W_C|}$ accounts for the (normalized) difference between the areas enclosed by the two cycles. In this geometrical context the losses are related to a comparison of areas which do not generate a real dissipation; this will be shown specifically in the next section (for an example, see Fig. 3a). Alternatively, as it will be seen in section 4, $d_e$ can be interpreted  in terms of the so-called endoreversibility~\cite{durmayaz04} since it represents an internally reversible cycle coupled to external heat baths. In this case the losses accounted for $d_e$  should be attributed to the coupling of the inner part with the external heat baths and the corresponding entropy-producing irreversibilities variation. The quantitative equivalence of both interpretations has been pointed out in~\cite{julian14}, where the role of the dissipation of an endoreversible engine in the context of finite-time thermodynamics (FTT) has been analyzed.

On the other hand, $d_i$ measures the differences in work output between the reversible and irreversible performances of $M$ and its origin is due to internal irreversibilities. We can say that $d_i$ is operational in nature since it accounts for internal irreversibilities (as frictions, viscosities and so on) in a real engine with respect to its idealized, reversible counterpart

\section{Brayton-like heat engine}\label{s3}
In order to show how the loss factors work and their influence on the power output and efficeincy of a heat engine we elect as a first representative example a simple irreversible Brayton heat engine, which is well known both in classical equilibrium thermodynamics and in more advanced disciplines devoted to optimization and control of energy conversion processes. The Brayton cycle is the usual model of a continuous-combustion gas turbine engine where compressed air enters in a combustion chamber and it is mixed with fuel. The mixture is burned out and the resulting gas at high temperature and high pressure is expanded across the turbine, where work is done and the gas returns to approximately the atmospheric pressure. Then, gas is exhausted to the ambient. 

\begin{figure}[t]\centering
\includegraphics[width=.49\linewidth]{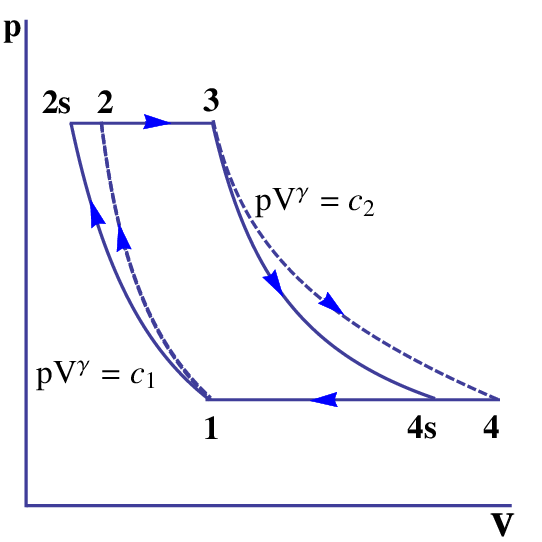}
\includegraphics[width=.49\linewidth]{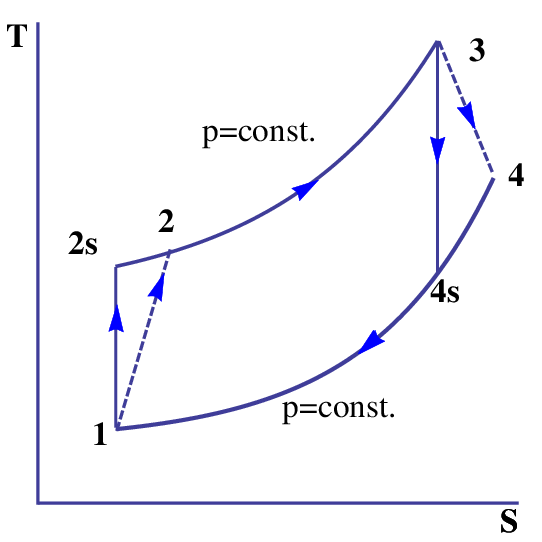}
\caption{Brayton-like cycle in the $p$-$V$ and $T$-$S$ diagrams. Irreversibilities are introduced in the adiabatic processes $1\rightarrow 2$ and $3\rightarrow  4$ (see text).}
\label{fig.2}
\end{figure}

Main internal irreversibilities in both recuperative and non-recuperative Brayton heat engine includes effects coming from fluid friction of air in the turbine and compressor. Volume increases in the heat addition and drops in the heat rejection processes~\cite{wylen94,hernandez95,gordon92}. Here and for simplicity we avoid the effects of the regenerator and the pressure drops while the irreversibilities in the compressor and turbine are accounted by the corresponding isentropic efficiencies $\eta_{c}$ and $\eta_{t}$, which quantify the degree to which adiabatic branches deviate from being isentropic. We assume a mass flow rate $\dot m$ as an ideal gas with constant heat capacities and running an ideal air-standard Brayton cycle $1\rightarrow 2\rightarrow 3\rightarrow 4$ (see Fig.~2). Then $\eta_{c}=\frac{T_{2s}-T_1}{T_{2}-T_1}\leq 1$ and $\eta_{t}=\frac{T_{3}-T_{4}}{T_{3}-T_{4s}}\leq 1$. Now $|\dot Q_{in}|=\dot m c_p(T_3-T_2)$ and $|\dot Q_{out}|=\dot m c_p(T_{4}-T_1)$. These magnitudes can be easily worked out in terms of the main characteristics of the cycle as $a=r_p^{\frac{\gamma-1}{\gamma}}$ ($r_p=\frac{P_{2s}}{P_1}=\frac{P_{3}}{P_4s}$), $\gamma=c_p/c_v$, $\tau=\frac{T_1}{T_3}$, and the isentropic efficiencies $\eta_{c}$ and $\eta_{t}$. The final results read as~\cite{wylen94,hernandez95,gordon92}:

\begin{figure}\centering
\includegraphics[width=.9\linewidth]{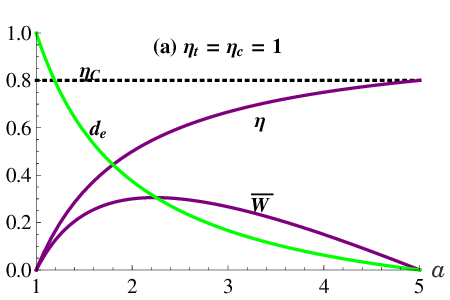}\\
\includegraphics[width=.9\linewidth]{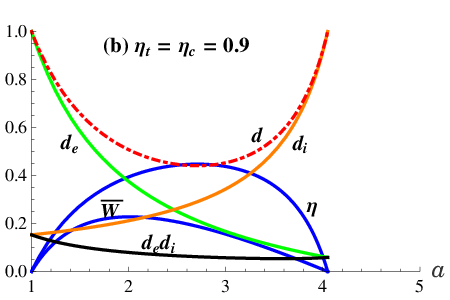}\\
\includegraphics[width=.9\linewidth]{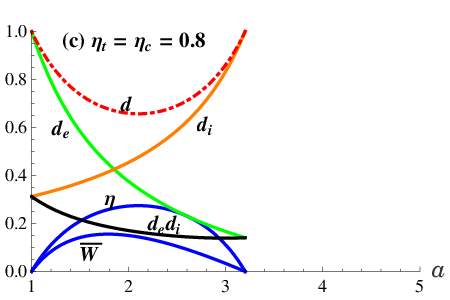}\\
\includegraphics[width=.9\linewidth]{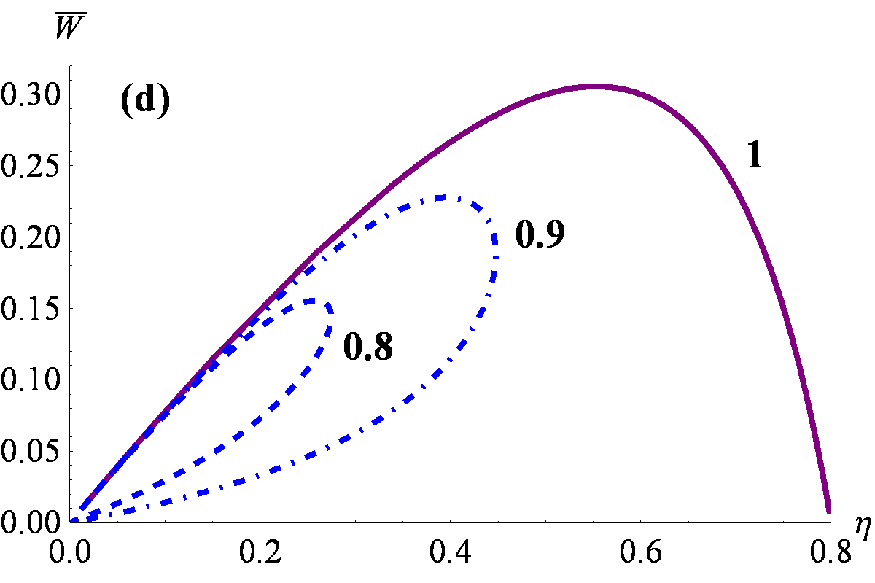}
\caption{Behavior of $\overline {W}(a,\tau,\eta_{c},\eta_{t})$, $\eta(a,\tau,\eta_{c},\eta_{t})$, $d_e(a,\tau)$, $d_i(a,\tau,\eta_{c},\eta_{t})$ and $d(a,\tau,\eta_{c},\eta_{t})$ for a Brayton cycle; see Eqs. (7)-(11). In (a) it is shown the behavior of $d_e$, $\eta$ and $\overline {W}$
for the reversible performance ($\eta_c=\eta_t=1$; that is, $d_i=0$). In (b) and (c) the evolution of $d_i$, $d_ed_i$, and $d$ is also plotted for values $\eta_c=\eta_t=0.9$ and $\eta_c=\eta_t=0.8$, respectively. In (d) the parabolic and loop behavior of the work-efficiency parametric curves is depicted. In all cases $\tau=0.2$ ($\eta_C=0.8$).}
\label{fig.3}
\end{figure}

\begin{eqnarray}
|\dot Q_{in}|(a,\tau,\eta_{c})=\dot m c_p(T_3-T_2)=\nonumber\\
\dot m c_pT_3\left[1-\tau-\frac{\tau }{\eta_{c}}(a-1)\right] \end{eqnarray}
\begin{equation}
\dot W(a,\tau,\eta_{c},\eta_{t})=\dot m c_pT_3\left[\eta_t-\frac{\tau a}{\eta_{c}}\right] \left
[1-\frac{1}{a}\right]
\end{equation}
\begin{equation}
\eta(a,\tau,\eta_{c},\eta_{t})=\frac{\left(\eta_t-\frac{\tau a}{\eta_{c}}\right)\left
(1-\frac{1}{a}\right)}
{1-\tau-\frac{\tau }{\eta_{c}}(a-1)}
\end{equation}
Note that when $\eta_{c}=\eta_{t}=1$, i.e., compressor and turbine internal irreversibilities are absent, Eqs.~(6)-(8) reduce to the well-known equations of a reversible Brayton cycle. In particular, the reversible efficiency is
\begin{equation}
\eta_{rev}(a,\tau)=\eta(a,\tau,\eta_{c}=1,\eta_{t}=1)=1-\frac{1}{a}
\end{equation}
From Eqs. (8) and (9), the loss factors in Eqs. (4) and (5) are given, respectively,  by:
\begin{equation}
d_e(a,\tau)=1-\frac{\eta_{rev}(a,\tau)}{1-\tau}=\frac{1-a\tau}{a(1-\tau)}
\end{equation}
\begin{eqnarray}
d_{i}(a,\tau,\eta_{c},\eta_{t})=1-\frac{\eta (a,\tau,\eta_{c},\eta_{t})}{\eta_{rev}(a,\tau)}=\nonumber\\
=\frac{\eta_c(1-\tau)+\tau-\eta_c\eta_t}{\eta_c(1-\tau)+\tau-a\tau}
\end{eqnarray}

In Fig.~3a we plot the behavior of normalized work output, $\overline {W}(a,\tau)\equiv \dot W(a,\tau)/\dot m c_pT_3$, the efficiency, and the external loss factor $d_e$ for the idealized version of the Brayton heat engine $\eta_{c}=\eta_{t}=1$ ($d_i\equiv 0$). In Figs. 3(b)-3(c) we show the evolution of the normalized power output $\overline{{W}}(a,\tau,\eta_{c},\eta_{t})\equiv \dot W(a,\tau,\eta_{c},\eta_{t})/\dot m c_pT_3$, the efficiency, the operating  loss factor $d_i$, the coupling term $d_ed_i$, and the total loss factor $d$ for values of $\eta_{c}=\eta_{t}=0.9$ and $0.8$, which account for increasing irreversibilities.

In the reversible performance, Fig.~3a, we observe how the efficiency monotonically increases from zero (at $a=1$) to the maximum possible Carnot value $\eta_C=1-\tau=0.8$ (at $a=1/\tau=5; \tau=0.2$) while $d_e$, which measures departures from the Carnot efficiency, shows a monotonically decreasing behavior from $1$ to $0$.
The work output shows a maximum value at $a=1/\sqrt{\tau}$ and two null values at the extreme allowed values of $a$: $a=1$ and $a=1/\tau$. A peculiar value is the efficiency at maximum work which take the values $\eta_{rev}(a=\frac{1}{\sqrt{\tau}},\tau)=1-\sqrt{\tau}\equiv \eta_{CA}$, \textit{i.e.}, the Curzon-Ahlborn efficiency~\cite{curzon75,hernandez15}. The parametric plot work-efficiency is an open parabolic-like curve (see Fig. 3d), characteristic of the so-called endoreversible cycles (internally reversible but with thermal couplings to external heat baths) and also of the most idealized cycles considered in classical equilibrium thermodynamics (Otto, Diesel, Atkinson). However, only those with two adiabatic branches alternating with two others of the same nature (Brayton, Otto) show the CA-efficiency at maximum work while those with two adiabatic branches alternating with two others of the different nature (Diesel, Atkinson) show an efficiency at maximum power depending on the specific physical properties of the working fluid, though nearly to the CA-value~\cite{leff87,gordon91}. 

The evolution with irreversibilities is shown in Figs.~3b and 3c. Opposite to $d_e$, the internal loss factor $d_{i}$ shows a monotonically increase with the pressure ratio $a$ since, as expected, higher pressure ratios induces higher internal losses in the compressor and turbine. The total loss factor $d=d_e+d_i-d_ed_i$ (see Eq.~3) thus shows an inverted parabolic-like shape modulated by the coupling term with a well defined minimum value at the same $a$-value as the maximum of the efficiency (and closer to maximum power) and with two maximum values ($d=1$) when efficiency (and power) are null. Indeed, as the frictional fluid losses increase (lower values of $\eta_{t}$ and $\eta_{c}$) $d_i$ (and $d_ed_i$) increases in intensity; as a direct consequence we observe lower values of the efficiency accomplished with lower values of power output. The evolution of the parametric plot power-efficiency is now loop-like (see Fig.~3d) with close but non-coincident maximum power and maximum efficiency points. Such behavior has been identified as a sign of actual heat engine models~\cite{gordon92,gordon91,qin03}. As internal irreversibilities increase the loop-like behaviors are narrower and with smaller maximum power and maximum efficiency values. 

\section{Irreversible Carnot-like heat engine}\label{s4}
\begin{figure}[t]\centering
\includegraphics[width=\linewidth]{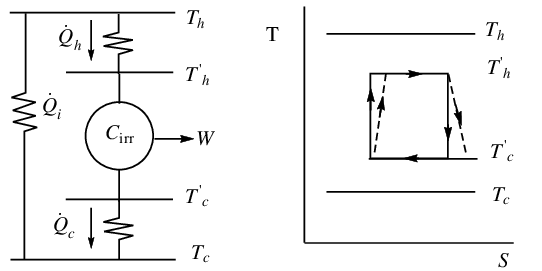}
\caption{Scheme of an irreversible Carnot-like heat engine (left) and its T-S diagram (right).}
\label{fig.4}
\end{figure}

As a second and representative example, we elect an irreversible Carnot-like heat engine. This model, as an extension of the original Curzon-Ahlborn endoreversible heat engine and related simplified versions~\cite{agrawal09, pescetti14}, has been widely used in finite-time thermodynamics~\cite{durmayaz04}. Following~\cite{jchen94} we assume a working system running a Carnot-like cycle where the isothermal heat addition and heat rejection processes are at temperatures $T'_h$  and $T'_c$, respectively, see Fig.~4. The inner part is coupled to external heat baths at temperatures $T_h$ and $T_c$ by appropriate heat transfer laws, $Q_h$ and $Q_c$, and heat leak $Q_i$ between the external reservoirs  is allowed. The simplest mathematical version corresponds to the assumption that all heat transfer processes follow a linear law with the temperature difference and constant thermal conductances $\sigma$. In this case the heat fluxes can be mathematically described by
\begin{equation}
{|\dot Q_h|} = \sigma_{\rm h}\left( T_{\rm h}-T'_{\rm h}\right)\equiv
\sigma_{\rm h}T_h \left(1-\frac{1}{a_h} \right)
\label{eq.qh}
\end{equation}
\begin{equation}
{|\dot Q_c|} = \sigma_{\rm c}\left( T'_{\rm c}-T_{\rm c}\right)\equiv
\sigma_{\rm c}T_c \left(a_c-1 \right)
\label{eq.qc}
\end{equation}
\begin{equation}
{|\dot Q_i|} = \sigma_{\rm i}\left( T_{\rm h}-T_{\rm c}\right)\equiv
\sigma_{\rm i}T_h \left(1-\tau \right)
\label{eq.qi}
\end{equation}
where $a_h=T_h/T'_h \geq 1$, $a_c=T'_c/T_c \geq 1$.

The reversible or irreversible nature of the inner cycle is accounted by the Clausius relation
\begin{equation}
I\frac{|\dot Q_c|}{T'_c}-\frac{|\dot Q_h|}{T'_h}= 0.
\label{eq.icl}
\end{equation}
If the inner cycle is reversible $I=1$ and if irreversible $I<1$. The two internal temperatures $T'_h$ and $T'_c$ (i.e., $a_h$ and $a_c$)  are not independent variables. Substituting  Eqs.~(\ref{eq.qh}) and ~(\ref{eq.qc}) in~(\ref{eq.icl}) we obtain:
\begin{equation}
a_c= \frac{I}{I-\sigma_{hc}(a_h-1)}
\label{eq.ic}
\end{equation}
where $\sigma_{hc}=\sigma_h/\sigma_c$. From the above equations, the power $\dot {|W|}=|\dot Q_h|-|\dot Q_c|$  and the efficiency $\eta=\dot {|W|}/(|\dot Q_h|+|\dot Q_i|)$ read as:

\begin{eqnarray}
\dot {|W|}(a_h, I, \sigma_h, \sigma_c, \tau)=\nonumber\\
\sigma_h T_h\frac{I\left(a_h-1\right)-\sigma_{hc}\left(a_h-1\right)^2-\tau \left(a_h^2-a_h\right)}
{a_h\left(I+ \sigma_{hc}\right)-\sigma_{hc} a_h^2}
\label{eq.power}
\end{eqnarray}
\begin{eqnarray}
\eta(a_h, I, \sigma_{hc}, \sigma_{ih},\tau)=\nonumber\\
\left[1-\frac{a_h \tau}{I-\sigma_{hc} \left(a_h-1\right)}\right]
\left[\frac{\left(a_h-1\right)}{\left(a_h-1\right)+\sigma_{ih}\left(1-\tau\right) a_h}
\right]
\label{eq.ren}
\end{eqnarray}
where  $\sigma_{ih}=\sigma_i/\sigma_h$. If $I=1$ and $\sigma_{i}= 0$ the model recovers the original Curzon-Ahlborn endoreversible model (as an internally reversible Carnot cycle externally coupled to external heat baths by linear heat transfer laws)\cite{callen85,curzon75}. 

The consideration of values $I<1$ and $\sigma_{i}\neq 0$ thus give additional information on the heat leak and internal dissipations. As in the previous section, the two loss factors are defined as:
\begin{equation}
d_e(a_h, \sigma_{hc},\tau)=1-\frac{\eta(a_h, I=1, \sigma_{hc}, \sigma_{ih}=0,\tau)}{1-\tau}
\end{equation}
\begin{equation}
d_{i}(a_h, I,\sigma_{hc}, \sigma_{ih}, \tau)=1-\frac{\eta (a_h, I, \sigma_{hc}, \sigma_{ih}, \tau)}{\eta(a_h, I=1, \sigma_{hc}, \sigma_{ih}=0,\tau)}
\end{equation}

\begin{figure}\centering
\includegraphics[width=.9\linewidth]{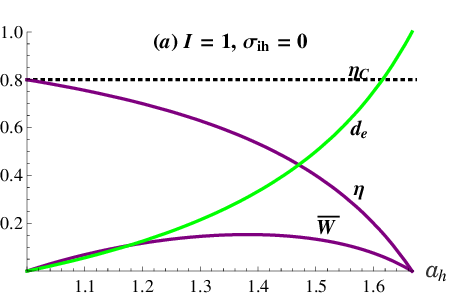}\\
\includegraphics[width=.9\linewidth]{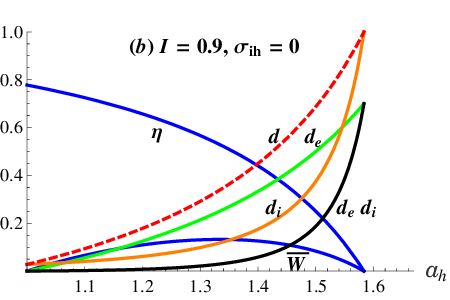}\\
\includegraphics[width=.9\linewidth]{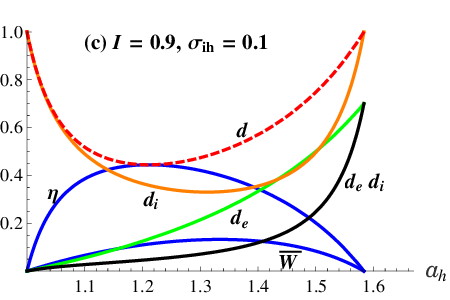}\\
\includegraphics[width=.98\linewidth]{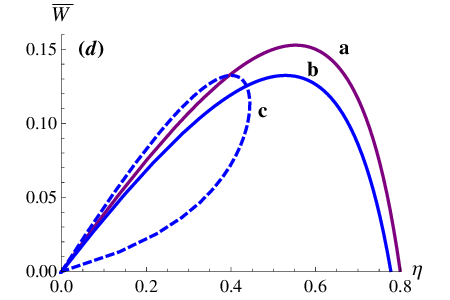}
\caption{Evolution of the behavior of  $\eta$  and normalized power $\overline W=\dot W/\sigma_hT_h$ (Eqs. 17 and 18) and $d_e$, $d_i$, $d_ed_i$ and $d$ (see Eqs. 19 and 20) for the irreversible Carnot-like heat engine for the labeled parameter $I$ and $\sigma_{ih}$. In (d) the corresponding parametric plots of the $\overline W-\eta$ curves are shown. In all cases $\sigma_{hc}=1$ and $\tau=0.2$.}
\label{fig.5}
\end{figure}

In the endoreversible limit ($I=1$, $\sigma_{ih}= 0$) the monotonic decrease of the efficiency $\eta$ is accomplished by the corresponding monotonic increase of the external loss factor $d_e$, see Fig.~5a. 
The influence of $I$ shows the same behavior for the internal loss factor $d_i$, see Fig.~5b.
As the influence of the heat leak ($\sigma_{ih}\neq 0$)  is taken into account we observe how $d_i$, $d_ed_i$ play a central role yielding in $d$ an inverted parabolic behavior, see Fig.~5c. As in above section for the Brayton cycle, the minimum of $d$ is coincident with the maximum efficiency and closer to the maximum power.
The resulting parametric power-efficiency curves are also shown in Fig.~5d.
Obviously, particular differences in $d_e$, $d_i$, $d_ed_i$ and $d$ between this model and the Brayton model arise due to the different nature of the external couplings (isobaric heat transfers in a Brayton heat engine or isothermal heat transfers in an endoreversible Carnot) and of internal dissipation (non adiabatic compressor and turbine performances in a Brayton cycle or heat leak and internal dissipation in the irreversible Carnot-like engine).

\section{Refrigeration cycles}\label{s5}

\begin{figure}\centering
\includegraphics[width=\linewidth]{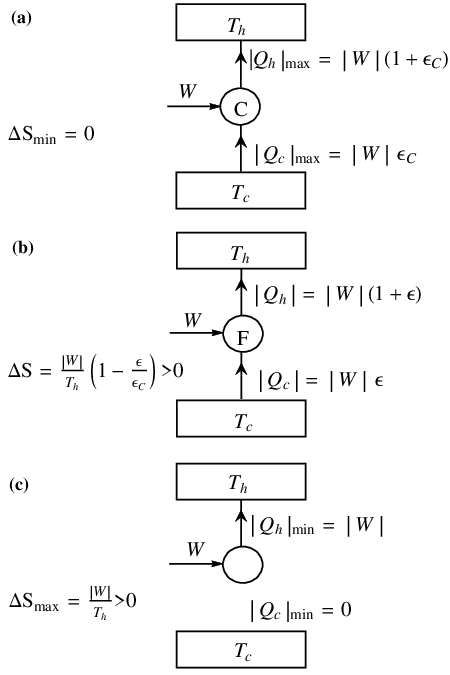}
\caption{Schematic diagram of a Carnot refrigeration cycle (a), a generic refrigerator (b) and of a direct dissipative process of work input in heat (c) between two external reservoirs at temperatures $T_h$ and $T_c$ (see text). Cooling load decreases from (a) to (c).}
\label{fig.8}
\end{figure}

The analysis for heat engine cycles in Sect.~II can be extended for refrigerator cycles. In this case our interest is the heat $Q_c\geq0$ extracted for the working fluid from a thermal reservoir at $T_c$ (the cold space). The second law states that the maximum coefficient of performance $\epsilon=Q_c/W$ is obtained when the working fluid runs a Carnot refrigeration cycle by the input of a work $W\geq0$, see Fig.~6a. Then $\epsilon_C=\frac{\tau}{1-\tau}$ and in this case the overall entropy production is null, $\Delta S=\frac{|Q_h|}{T_h}-\frac{|Q_c|}{T_c}=0$, and $Q_c\geq0$ is the maximum possible value given by $Q_c=W\epsilon_C=Q_{c,max}\equiv Q_{c,C}$. In the opposite side we have a situation, see Fig~6c, where with the same work input $W$ the cycle works between the same external temperatures without any heat extracted from the thermal bath at $T_c$. All input work is dissipated to the hot reservoir. In such a situation the entropy variation is the maximum possible $\Delta S=\frac{|Q_h|}{T_h}=\frac{W}{T_h}=\Delta S_{max}$, the COP is null and no heat is extracted from the low temperature reservoir, $Q_c=0$. In between these two extreme cases we consider a generic refrigerator F working between the same temperatures and with the same energy requirement $W$, see Fig.~6b. Now $Q_c=W\epsilon$, $\Delta S=\frac{Q_c+W}{T_h}-\frac{Q_c}{T_c}=\frac{W}{T_h}\left( 1-\frac{\epsilon}{\epsilon_C}\right)$ and the difference of heat absorbed in comparison with the most favorable case is $Q_{c,lost}=W(\epsilon_C-\epsilon)$. In order to account for losses we focus on the ratio between the lost refrigeration load to the maximum possible refrigeration load for the same work requirement:
\begin{equation}
d=\frac{Q_{c,lost}}{Q_{c, max}}=\frac{Q_{c,max}-Q_c}{Q_{c, max}}\equiv \frac{Q_{c,C}-Q_c}{Q_{c, C}}=1-\frac{\epsilon}{\epsilon_C}
\end{equation}
where the last equality follows from the assumption of equal work input in both cycles.
Following the same steps that for heat engines we divide deviations from the Carnot performance of any actual refrigerator in two contributions as:
\begin{equation}
d=\frac{\epsilon_C-\epsilon}{\epsilon_C}\equiv \frac{\epsilon_C-\epsilon_{rev}}{\epsilon_C}+\left (\frac{\epsilon_{rev}-\epsilon}{\epsilon_{rev}}\right ) \frac{\epsilon_{rev}}{\epsilon_C}\le 1
\end{equation}
or
\begin{equation}
d=d_e+\left(\frac{\epsilon_{rev}}{\epsilon_C}\right)d_i=d_e+d_{i}-d_e d_{i}\label{eq:dRE}
\end{equation}
with
\begin{equation}
d_e=1-\frac{\epsilon_{rev}}{\epsilon_C}\le 1
\end{equation}
and
\begin{equation}
d_{i}=1-\frac{\epsilon}{\epsilon_{rev}}\le 1
\end{equation}

The application of above definition to a particular cycle needs the specification of the reversible and irreversible models which are compared with the Carnot refrigeration between the same external reservoirs. 
Here we don't present any particular result but the evaluation of the loss factors for the inverted Brayton and Carnot-like cycles presented above could be an instructive exercise for students and interested readers. However, special care should be taken when a heat engine cycle is reversed~\cite{ferreira12}. Only the Carnot cycle automatically yields a refrigerator (heat pump) cycle in all circumstances. For other well-known cycles (as Otto, Diesel, Brayton, Atkinson)  the inversion can yield refrigerator cycles only under restricted conditions for the involved temperatures~\cite{ferreira12}.

\section{Some remarks and conclusions}\label{s7}

Notice that the form for counting losses expressed in Eq.~(\ref{eq:dHE})  for heat engines and Eq.~(\ref{eq:dRE}) for refrigerators is the same. In any case, when one of the loss factor is zero the total loss factor is due to the other one, as would be expected. However, an interesting feature of this expression is the negative crossing term in Eq.~ (\ref{eq:dHE}) or Eq.~(\ref{eq:dRE}) when both kind of losses are present. In this case the simple addition of losses give an over-counting which must be decreased by the interplay accounted by the negative term.
The interplay between the external and internal loss factors to account for the so called external and internal irreversibilities~\cite{wu15} could be of some utility to explain observed reversible landmarks for efficiencies stemming from finite-time thermodynamics~\cite{julian12} as the maximum work efficiency of some reversible cycle. From the analyzed examples we see that loop-like behavior of the power-efficiency in heat engines curves can be attributed in general to the internal irreversibilities while the open, parabolic-like behavior comes from the external losses. Accordingly,  
the interplay between both $d_e$ and $d_i$ being non zero accounts for the participation of both quantities in the appearing of such behavior.\\

Recently, Wu~\cite{wu15} published an instructive paper on the three factors causing the thermal efficiency of a heat engine to be less than unity. The first factor is the nonzero reference point due to the unreachable absolute zero temperature and, as a consequence, is induced by unavailability rather than irreversibility. The second factor is the external irreversibility coming from coupling of the heat engine to external heat sources; and the third one is coming from the inherent internal irreversibility. From a quantitative point of view Wu quantifies the second-law overall efficiency of the heat engine $\eta_{II}=\eta_I/\eta_C$ \cite{bejan06,marcella92} as $\eta_{II}=\eta_{int,ir}\eta_{ext,ir}$, where $\eta_{ext,ir}=\eta_{rev}/\eta_C$ and $\eta_{int,ir}=\eta_{I}/\eta_{rev}$. In these expressions $\eta_{rev}$ is the efficiency of ideal cycles as those which are internally reversible but they may involve irreversibilities that are external to the system, $\eta_C$ is the Carnot efficiency of the Carnot cycle working between the same extreme temperatures, and $\eta_I$ is the first-law thermal efficiency. However, one must be careful when considering the second factor, since an external coupling with two reservoirs is a characteristic of a endoreversible model rather than that of a reversible system. In other words, it must be specified if the irreversibility is real (intrinsic to the model) or coming from a geometrical nature (no real dissipation).
Note that in our analysis the total loss factor is $d=1-\eta_{II}$ while $d_e=1-\eta_{ext,ir}$ and $d_i=1-\eta_{int,ir}$. For refrigerator cycles, not considered by Wu~\cite{wu15}, the similarity can be extended straightforwardly: $d=1-\epsilon_{II}$ where $\epsilon_{II}$ is the second law COP defined ~\cite{bejan06,marcella92} as the ratio of the firs-law COP $\epsilon_I$ to the Carnot COP $\epsilon_C$, $\epsilon_{II}=\epsilon_I/\epsilon_C$.\\

In summary, the simple and pedagogical analysis presented here allows for a clear visualization of two classes of losses in the specific job of a heat device when measured from the corresponding Carnot performance. The first one measures the losses due to external coupling of the idealized, reversible operation with external heat baths while the second one measures losses of the actual performance coming from internal dissipations. The analysis does not need explicit evaluation of the entropy generation 
to link the thermodynamic non-ideality of the design to the physical characteristics of the system,  a subtle and sometimes difficult task.
The results could guide students and readers to a more detailed knowledge on how external and internal losses, together with their interplay, affect the performance of real heat devices. Hence, it is a starting point for more advance studies on modeling and optimization of real devices and installations, an unavoidable task in a sustainable use of the natural resources.
\subsection*{Acknowledgment}
Angulo-Brown F an Gonzalez-Ayala J acknowledge CONACYT-M{\'E}XICO and COFFA-SIP-IPN-M\'EXICO; Velasco S and Calvo Hern\'andez A acknowledge Ministerio de Econom{\'i}a y Competitividad of Spain under Grant ENE2013-40644R.

\end{document}